# A Carpet Cloak Device for Visible Light

*Majid Gharghi[1a], Christopher Gladden[1a], Thomas Zentgraf[2], Yongmin Liu[1], Xiaobo Yin[1], Jason Valentine[3], Xiang Zhang[1,4,*]*

[1]NSF Nanoscale Science and Engineering Center (NSEC), University of California, Berkeley
3112 Etcheverry Hall, UC Berkeley, CA 94720, USA
[2]Department of Physics, University of Paderborn
Warburger Str. 100, 33098 Paderborn, Germany
[3]Department of Mechanical Engineering, Vanderbilt University
VU Station B 351592, Nashville, TN 37235, USA
[4]Materials Science Division, Lawrence Berkeley National Laboratory,
1 Cyclotron Road, Berkeley, CA 94720, USA

[a] These authors contributed equally to this work.

*To whom correspondence should be addressed. E-mail: xiang@berkeley.edu

**ABSTRACT:**

We report an invisibility carpet cloak device, which is capable of making an object undetectable by visible light. The cloak is designed using quasi conformal mapping and is fabricated in a silicon nitride waveguide on a specially developed nano-porous silicon oxide substrate with a very low refractive index. The spatial index variation is realized by etching holes of various sizes in the nitride layer at deep subwavelength scale creating a local effective medium index. The fabricated device demonstrates wideband invisibility throughout the visible spectrum with low loss. This silicon nitride on low index substrate can also be a general scheme for implementation of transformation optical devices at visible frequency.

Invisibility cloaks, a family of optical illusion devices that route electromagnetic (EM) waves around an object so that the existence of the object does not perturb light propagation, are still in their infancy. Artificially engineered materials with specific EM properties, known as metamaterials [1,2], have been used to control the propagation of EM waves, and have recently been applied to cloaking through transformation optics [3-8]. The invariance of Maxwell's equations under optical coordinate transformation allows the space around the object to be reshaped such that the light can propagate in the desired way. Such transformations usually require EM properties with extreme values that are only achievable in metallic metamaterials, and have been experimentally demonstrated for cloaking in microwave frequencies [9,10]. Due to the significant metallic loss at optical frequencies, the implementation of such cloaks for visible light has been difficult. Recently another innovative strategy was developed based on exploiting uniaxial crystals [11,12]. These devices have demonstrated cloaking in visible frequencies for a certain polarization of light based on intrinsic anisotropy in the crystals. As an alternative, conformal mapping, where an inverse transformation of the electrical permittivity and magnetic permeability leads to a spatially variable refractive index profile [13], can be applied to isotropic dielectric metamaterials. While 3D conformal mapping leads to anisotropic index profiles [14], a 2D quasi conformal mapping (QCM) can be employed to minimize anisotropy. The 2D QCM is the basis for the carpet cloak [15], where the object is hidden under a reflective layer (the carpet). To achieve cloaking, the raised protrusion (the bump) created in the reflective layer is mapped to a flat plane and the resulting 2D index profile forms a carpet cloak device. In contrast to resonant optical structures [16,17], QCM carpet cloak provides a broadband loss-less design and may be invariably extended in the third direction with some limitations [18], experimentally demonstrated to operate for a range of viewing angles [19]. The relatively modest materials requirement from QCM enabled the implementation of the cloaking devices in the infrared [20,21] using a silicon waveguide. The index variation in these systems is realized through a spatial modulation of the filling fraction of dielectrics at subwavelength dimensions, providing a weighted average index according to effective medium approximation. At shorter wavelengths however, devices reported so far suffer from significant scattering from surfaces within a unit cell as the feature sizes become comparable to wavelength. More importantly, the silicon device layer employed in the infrared becomes lossy due to absorption at visible frequencies. To realize cloaking at visible frequencies, the unit cell size must be reduced and a new materials system is required that provides both transparency and sufficient index contrast. We demonstrate here a visible light carpet cloak device made of a dielectric silicon nitride on a specially prepared nano-porous silicon oxide with very low index of ~1.2. This unique substrate increases the available index modulation and enables the implementation of transformation optics for visible guided light.

The optical transformation is designed so that a bump centered at the origin and defined analytically as $y = h \cdot \cos^2(\pi \frac{x}{w})$ with a height $h$ of 300nm and a width $w$ of 6μm is compressed in y direction (inset of Figure 1a). QCM of the transformed space results in the relative index variation shown in Figure 1a. The smallest index values occur at the corners of the bump and maximum index appears around the top of the curved region. Away from the bump, the index converges to the relative background index of 1. The actual index variation Δn and the background index in the final device depend on the materials used. Figure 1b shows the index modulation required to achieve cloaking and the substrate index for the waveguide, highlighting the requirements to implement the carpet cloak in the visible spectrum. It is important to note that silicon oxide substrate index fails to provide enough contrast for waveguiding at visible frequencies. To accomplish the required index contrast, we consider a silicon nitride (SiN) waveguide on a specially developed low index substrate made from nano-porous silicon oxide. The inset in Figure 1b shows the approximate unit cell size requirement for the scattering-free operation of the effective medium. In order for the device to operate throughout the visible spectrum, feature sizes should be on the order of 65nm half pitch.

In visible frequencies, our SiN has a bulk refractive index of approximately 1.9 and the nano-porous silicon oxide substrate has an exceptionally low refractive index of ~1.2 and both are transparent. To modulate the index of the SiN waveguide we drill variable sized holes in a 2D hexagonal lattice with 130nm pitch, thereby changing the filling fraction of air/SiN mixture. A waveguide thickness of 300nm is chosen to achieve single mode propagation while maintaining sufficient index contrast for strongly confined modes at longer wavelengths. This will result in the fundamental transverse electric mode index varying from 1.83 to 1.74 for wavelengths of 400nm to 700nm. Thus the background index calculated for the transformation in Figure 1a was projected to approximately 1.5. Figure 2a shows a cross-sectional schematic of the device with the hole size pattern generated to achieve the required index modulation. As expected, the largest holes are located at the corners of the bump and the smallest holes appear right above the peak of the bump. Additionally, an area of constant hole size surrounding the cloak device is required to provide a uniform background index.

The devices fabrication process begins with the preparation of the low index substrate. A crystalline silicon wafer is electrochemically etched in an acid/organic (1:1 hydrofluoric acid/ethanol) solution to produce a porous network that penetrates several micrometers into the top surface. The porosity of this network is increased by oxidizing several atomic layers of silicon on the internal surfaces at 300°C and subsequently removing this oxide layer by selective etching. By repeating the oxidation and etching several times, the solid silicon is slowly consumed, leaving solid filling fractions as low as 15%. Once the desired porosity is reached, the whole silicon network is converted to a porous silicon oxide medium by oxidizing at high temperature (800°C). Due to expansion during oxidation, the resulting medium turns into an approximately 65% porous glass, where the pores range in size from 2nm to 20nm (shown in the inset of Figure 2a). The resulting substrate has a refractive index below 1.25 as confirmed by optical interference measurement at infrared, and has a smooth surface with roughness less than 3nm RMS. A 300nm SiN slab waveguide is deposited on the low index substrate using plasma enhanced chemical vapor deposition (PECVD).

To fabricate the hole pattern introduced in Figure 2a, we use a two step pattern transfer process. The waveguide is first covered with a 100 nm thick PECVD silicon oxide to act as a hard mask for etching. ZEP520A electron beam resist is then spun on the oxide layer with a thickness of 120nm, and the generated hole pattern is written on the resist by electron beam lithography. Figure 2b shows an atomic force microscope (AFM) image of the developed resist with the hole sizes varying from 65nm to 20nm. The resist is used as a mask to transfer the pattern to the 100nm oxide layer by reactive ion etching (RIE). The resist is then ashed and the pattern is transferred to the nitride waveguide using a selective RIE process, followed by removal of the oxide mask. Figure 2c shows an overall scanning electron microscope (SEM) image of the fabricated structure including a triangular region defining background index. To achieve a reflective bottom plane for the device, the area under the bump is opened to the edge of the die using focused ion beam (FIB) milling. The structure is then mounted at 90° angle and a directional deposition is performed using electron beam evaporation to cover the bottom of the bump with silver.

The optical characterization setup is depicted in Figure 3a. For exciting optical modes in the waveguide, we use a pulsed femto-second (FS) Ti-Sapphire laser (Spectra Physics MaiTai) along with an optical parametric oscillator (Spectra Physics Inspire OPO) as the source. To couple light into and out of the waveguide, two gratings are made parallel to the sides of the triangle (inset Fig 3a). This design minimizes scattering from corners and keeps the propagation normal to all abrupt interfaces. The gratings consist of three periods to allow relatively broadband coupling across the visible spectrum. The laser beam was focused on the input grating through a 50x dark field objective, and the light collected from the output grating was imaged using a CCD camera (Figure 3c). Since the input beam scatters at the in-coupling grating, a polarizing filter aligned to the output grating was placed in front of the CCD to reduce the intensity of the image of the input beam while not affecting the output beam. The coupling and the propagation of the beam in the slab waveguide were checked using two parallel gratings (transmission mode) in all configurations. In an actual experiment, besides the non-perfect coupling at the gratings, there is a small additional loss due to the reflection at each of the triangle sides. However all such losses are extrinsic to the transformation device.

Two control samples were also fabricated beside the cloak sample, a bump structure without any transformation pattern and a simple mirror without a bump. To characterize the performance of the cloaking transformation, the reflection of the flat mirror, the non-cloaked bump, and the cloak device are shown in Figure 4 for three different excitation wavelengths of 480nm, 520nm, and 700nm. The 700nm beam is the direct output from the Ti-Sapphire laser, whereas the other two are the outputs of the second harmonic generation in the OPO, and thus the beam profiles are slightly different at each wavelength. The figure also shows the expected propagation pattern of an impinging Gaussian beam in the absence (cloak off) and the presence (cloak on) of the transformation as simulated for an example wavelength of 600nm. The reflection from the non-cloaked bump shows a clear perturbation in the wavefront when compared to the reflection from the flat mirror. The cloak device on the other hand, reconstructs the wavefront producing a beam profile identical to the original Gaussian beam reflected from the flat mirror. Since the gratings are not optimized for each particular wavelength, the reflected beam at 700nm scatters with a two-lobe pattern at the output grating irrespective of the light propagation and reflection inside the waveguide. This confirms that the designed transformation effectively cloaks the uneven surface throughout the entire visible spectrum, successfully rendering any object behind it invisible at far field. To further test the device, the beam was also scanned along the length of the input grating to ensure the beam profile on the output grating is not perturbed.

The proposed cloak structure is inherently wideband and low-loss. However, the fabricated device has some limitations. The nitride material becomes lossy due to absorption as we approach the ultraviolet region. Also, the waveguide confinement at longer wavelength reduces the mode index, creating a waveguide cut-off in the infrared. Although the waveguide test structure and coupling gratings were designed for demonstration of the cloaking effect using light with transverse electric polarization, in principal the QCM transformation operates for both polarizations [15,19,20].

In conclusion, we have demonstrated a full visible spectrum transformation optical device that is capable of cloaking any object underneath a reflective carpet layer. In contrast to the previous demonstrations that were limited to the infrared light, this work makes actual invisibility for the light seen by the human eye possible. In addition to successful demonstration of cloaking in the visible range, the fabrication scheme employed here is a significant step toward general implementation of optical transformation structures in the visible range. The considerable index contrast between the waveguide and the substrate enables realization of large index variations in the metamaterial device. In addition, the nitride waveguide on the low index substrate provides a platform for on-chip photonic devices in the visible range, which have been traditionally achievable in silicon waveguide on oxide for the infrared.

**ACKNOWLEDGMENT**

The authors acknowledge funding support from the US Army Research Office (MURI programme W911NF-09-1-0539) and the US National Science Foundation (NSF Nanoscale Science and Engineering Center CMMI-0751621). Majid Gharghi acknowledges fellowship from Natural Sciences and Engineering Research Council of Canada (NSERC). Chris Gladden acknowledges support from NSF Graduate Research Fellowship Program.

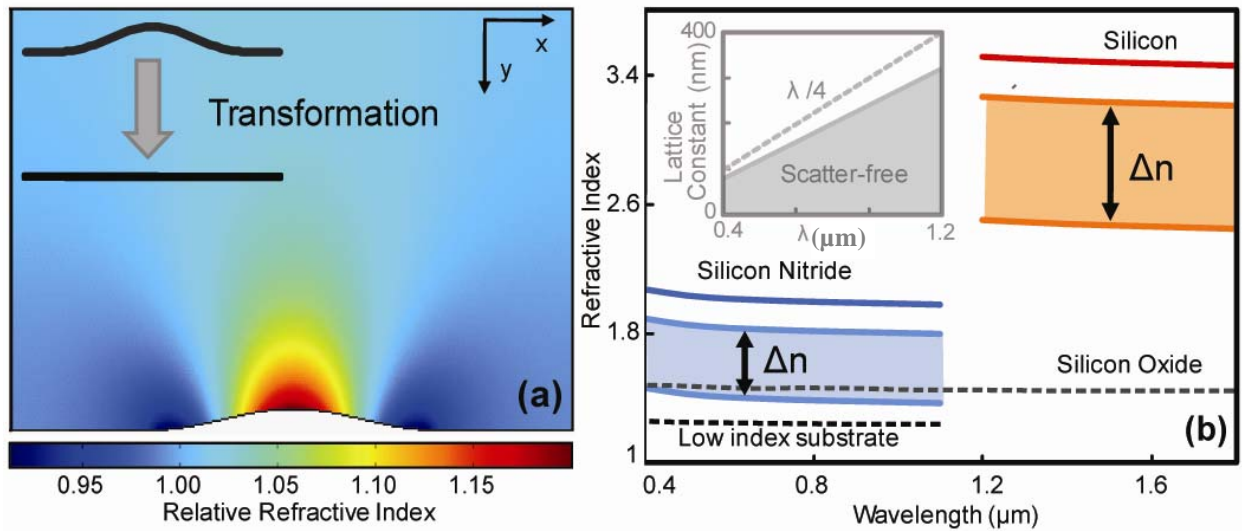

Figure 1: (a) Index variation for the optical transformation of a "bump" to a flat plane, normalized to index 1; the effective index of the cloak is found by multiplying the index shown by the desired background index; the inset schematically shows the transformation performed. (b) Requirements on the index variation and substrate index for implementing the cloaking transformation in the infrared (silicon waveguide on silicon oxide substrate) and the visible (silicon nitride waveguide on low index substrate). Solid line indicates the index of the unmodified material, while the shaded region below the solid line shows the range of index Δn that must be available; in visible frequencies a new substrate material is needed as the index variation required for a silicon nitride waveguide goes below the index of the conventional silicon oxide substrate. The inset shows the feature size requirement for scattering free propagation of light in the metamaterial as a function of wavelength, which is used as design criteria for the cloak.

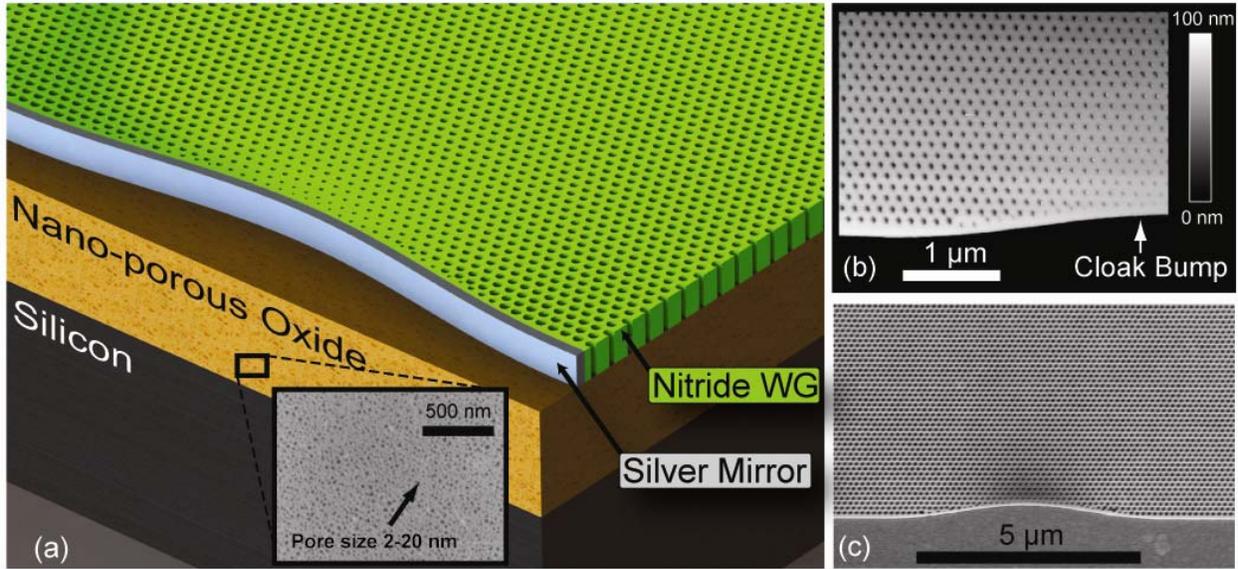

Figure 2: (a) Cross sectional schematic of the cloak device implemented in a silicon nitride waveguide on a low index nano-porous silicon oxide substrate; the nitride layer and the nano-porous oxide layer are 300nm and 5-10 μm thick respectively. The hole pattern allows for index modulation by varying the solid filling fraction. The holes vary in size from 65nm to 20nm. The inset shows an SEM image of the low index nano-porous silicon oxide substrate. (b) AFM image of the hole pattern as transferred to the electron beam resist after development; the smallest holes appear shallow due to incomplete penetration of the AFM tip, but based on the uniform pattern transfer we conclude that they are completely through the resist. (c) SEM image of the device, consisting of roughly 3000 holes; the optical transformation device is surrounded by a triangular background index region not pictured.

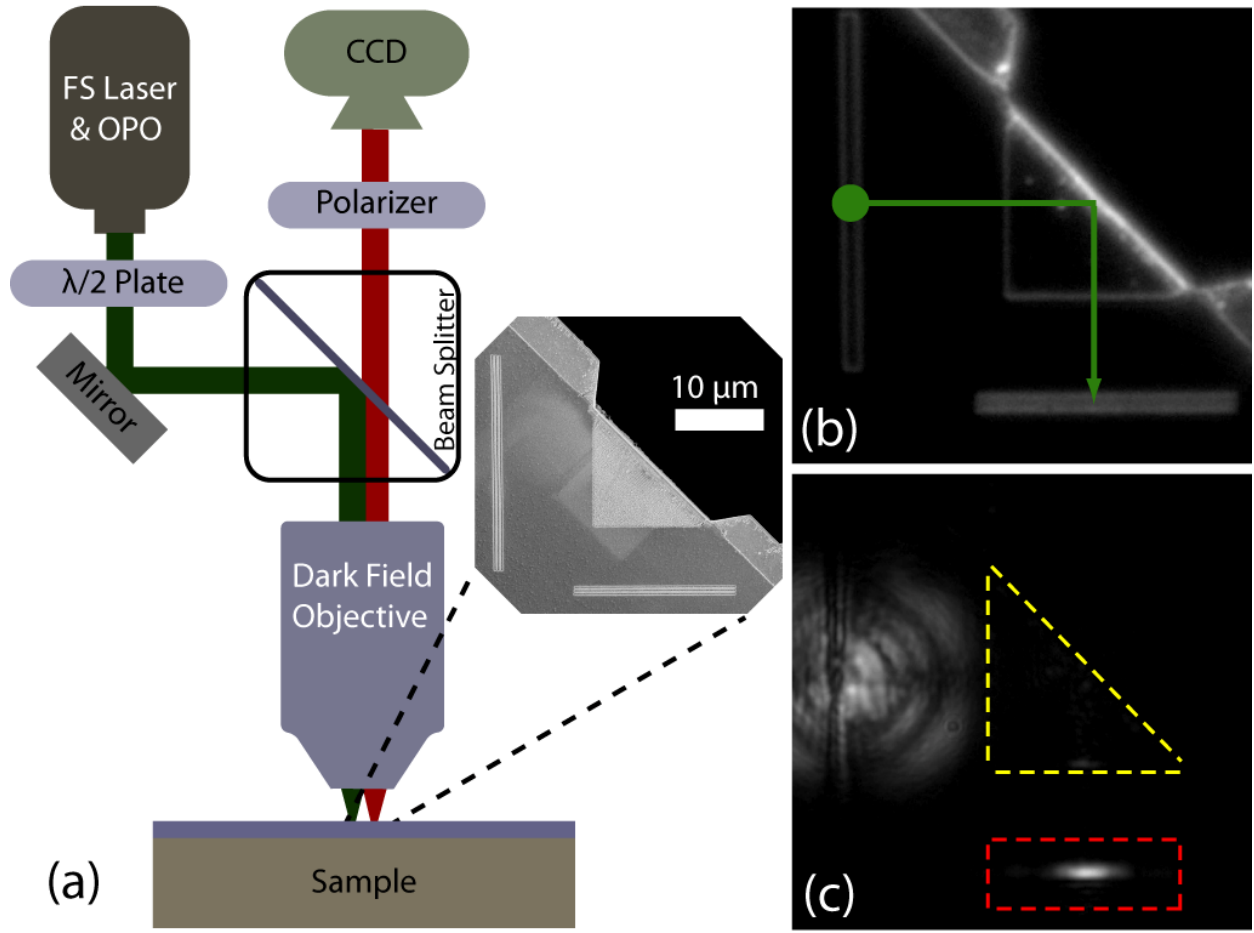

Figure 3: (a) Optical characterization setup; the half-wave plate rotates the polarization of the laser output as desired; the out-coupled light polarization is rotated 90° due to the reflection from the mirror in the sample, and the output polarizer is used to partially filter out the scattered input beam. Inset shows an SEM image of the gratings; the area behind the cloak has been etched by FIB and a silver layer has covered the back of the bump (b) Dark field image of the cloak device with the input and output gratings. The input grating and beam path are highlighted by the green circle and the green arrows respectively. (c) Dark field image of the in coupling laser and the out-coupled light. The red dashed rectangle indicates the out-coupling grating, which is the region of interest for the results presented in Figure 4.

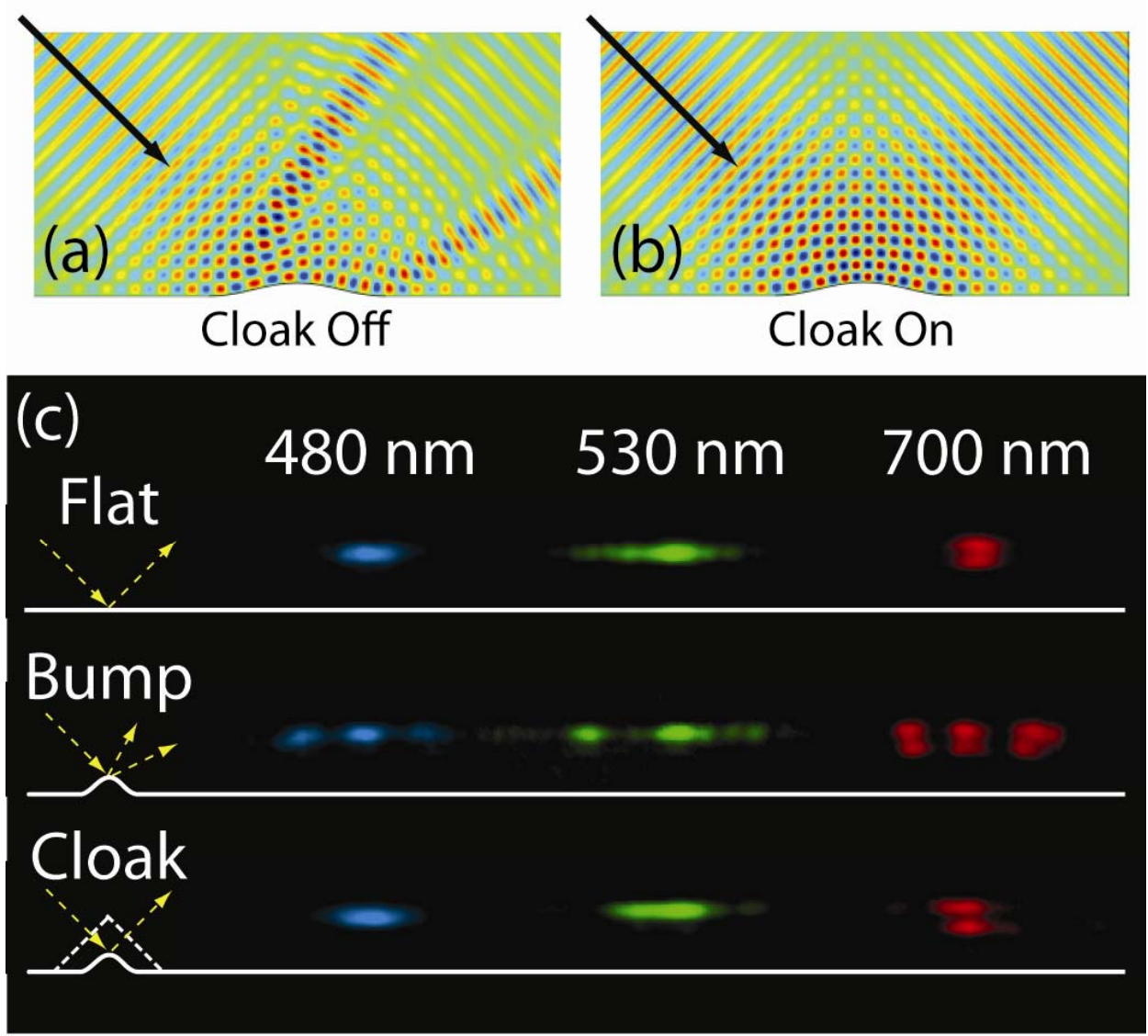

Figure 4: Simulated propagation of a Gaussian beam when reflected from a bumped carpet (a) without and (b) with a cloaking device on top of the carpet. The input beam is indicated by the black arrow has a wavelength of 600nm. The perturbation caused by the bump is masked by the cloak device and the reflected beam is reconstructed as if the bump did not exist. (c) Experimental results: normalized intensity CCD images of the out-coupled light (shown in false color) from the waveguide after reflection from a flat mirror, bump without a cloak, and cloaked bump at three different wavelengths. The bump clearly distorts the wavefront, creating a lobed pattern. The cloak successfully reconstructs the wavefront so that the output is unperturbed.

**REFERENCES**

(1) Pendry J. B.; Holden A. J.; Robbins D. J.; Stewart W. J. Magnetism from Conductors and Enhanced Non-Linear Phenomena, IEEE Trans. Microwave Theory Techniques 1999, 47, 2075-2084.

(2) Smith D. R.; Padilla W. J.; Vier D. C.; Nemat-Nasser S. C.; Schultz S. Composite Medium with Simultaneously Negative Permeability and Permittivity," Phys. Rev. Lett. 2000, 84, 4184-4187.

(3) Pendry J. B.; Schurig D.; Smith D. R. Controlling Electromagnetic Fields, Science 2006, 312, 1780-1782.

(4) Cai W.; Chettiar U. K.; Kildishev A. V.; Shalaev V.M. Optical Cloaking with Metatmaterials, Nature Phot. 2007, 1, 224-227.

(5) Rahm M.; Cummer S. A.; Shurig D.; Pendry J. B.; Smith D. R. Optical Design of Reflectionless Complex Media by Finite Embedded Coordinate Transformations, Phys. Rev. Lett. 2008, 100, 063903.

(6) Jiang W. X.; Cui T. J.; Yang X. M.; Cheng Q.; Liu R.; Smith D. R. Invisibility Cloak without Singularity, App. Phys. Lett. 2008, 93, 194102.

(7) Kildishev A. V.; Shalaev V. M. Engineering Space for Light via Transformation Optics, Opt. Lett. 2008, 33, 43-45.

(8) Chen H.; Ng J.; Lee C. W. J.; Lai Y.; Chan C. T. General Transformation for Reduced Invisibility Cloak, Phys. Rev. B 2009, 80, 085112.

(9) Schurig D.; Mock J. J.; Justice B. J.; Cummer S. A.; Pendry J. B.; Starr A. F.; Smith D. R. Metamaterial Electromagnetic Cloak at Microwave Frequencies, Science 2006, 314, 977-980.

(10) Edwards B.; Alu A.; Silveirinha M. G.; Engheta N. Experimental Verification of Plasmonic Cloaking at Microwave Frequencies with Metamaterials, Phys. Rev. Lett. 2009, 103, 153901.

(11)    Chen X.; Luo Y.; Zhang J.; Jiang K.; Pendry J. B.; Zhang S. Macroscopic Invisibility Cloaking of Visible Light, Nature Comm. 2011, 2, 176.

(12)    Zhang B.; Luo Y.; Liu X.; Barbastathis G. Macroscopic Invisibility Cloak for Visible Light, Phys. Rev. Lett. 2011, 106, 033901.

(13)    Leonhardt U. Optical Conformal Mapping, Science 2006, 312, 1777-1780.

(14)    Leonhardt U. Notes on Conformal Invisibility Devices, New J. Phys. 2006, 8, 118.

(15)    Li J.; Pendry J. B. Hiding under the carpet: A new strategy for cloaking, Phys. Rev. Lett. 2008, 101, 203901.

(16)    Alu A.; Engheta N. Multifrequency Optical Invisibility Cloak with Layered Plasmonic Shells, Phys. Rev. Lett. 2008, 100, 113901.

(17)    Smolyaninov I. I.; Smolyaninova V. N.; Kildishev A. V.; Shalaev V. M. Anisotropic Metamaterials Emulated by Tapered Waveguides: Application to Optical Cloaking, Phys. Rev. Lett. 2009, 102, 213901.

(18)    Landy N. I.; Kundtz N.; Smith D. R. Designing Three-Dimensional Transformation Optical Media Using Quasiconformal Coordinate Transformations, Phys. Rev. Lett. 2010, 105, 193902.

(19)    Ergin T.; Stenger N.; Brenner P.; Pendry J. B.; Wegener M. Three-dimensional invisibility cloak at optical wavelengths, Science 2010, 328, 337–339.

(20)    Valentine J.; Li J.; Zentgraf T.; Bartal G.; Zhang X. An optical cloak made of dielectrics, Nat. Mater. 2009, 8, 568–571.

(21)    Gabrielli L. H.; Cardenas J.; Poitras C. B.; Lipson M. Silicon nanostructure cloak operating at optical frequencies, Nat. Photonics 2009, 3, 461–463.